\documentclass{aa}

\usepackage{psfig}

\def\src  {SGR~1900$+$14}
\def\zerosei  {SGR~1806--20}
\def\integral {{\it INTEGRAL}}

\begin{document}



 \title{Magnetars as persistent hard X-ray sources:
 INTEGRAL discovery  of a hard tail in \src\thanks{Based on observations with INTEGRAL, an ESA project with instruments and the science data centre funded by ESA member states (especially the PI countries: Denmark, France, Germany, Italy, Switzerland, Spain), Czech Republic and Poland, and with the participation of Russia and the USA.} }

   \author{D. G\"{o}tz\inst{1}, S. Mereghetti\inst{2}, A. Tiengo\inst{2}, \and P. Esposito\inst{2,3}}

   \offprints{D. G\"{o}tz, email: diego.gotz@cea.fr}

   \institute{CEA Saclay, DSM/DAPNIA/Service d'Astrophysique, F-91191, Gif-sur-Yvette, France
   \and
   INAF -- Istituto di Astrofisica Spaziale e Fisica Cosmica,
                        Via Bassini 15, I-20133 Milano, Italy
    \and
    Dipartimento di Fisica Nucleare e Teorica -- Universit\`{a} degli Studi di Pavia, Via Bassi 6, I-27100 Pavia, Italy}


\abstract{Using 2.5 Ms of data obtained  by the \integral~ satellite
in 2003-2004, we discovered persistent hard X-ray  emission from
the soft gamma-ray repeater \src. Its 20-100 keV spectrum is well
described by a steep power law with photon index
$\Gamma$=3.1$\pm$0.5 and flux 1.5$\times$10$^{-11}$ erg cm$^{-2}$
s$^{-1}$.
Contrary to \zerosei, the only other soft gamma-ray repeater for
which persistent emission above 20 keV was reported, \src\ has
been detected in the hard X-ray range while it was in a quiescent
state (the last bursts from this source were observed in 2002).
By comparing the broad band spectra (1-100 keV) of all the
magnetars detected by \integral~ (the two SGRs and three anomalous
X-ray pulsars) we find evidence for a different spectral behaviour
of these two classes of sources.
 \keywords{gamma-rays: observations -- pulsars: individual \src} -- pulsars: general}
\authorrunning{D. G\"{o}tz et al.}
\titlerunning{Persistent hard X--ray emssion from \src}
\maketitle

\section{Introduction}

Soft gamma-ray repeaters (SGRs, for a recent review see
\cite{woodsrew}) are a small group of peculiar high-energy sources
generally interpreted as ``magnetars'', i.e. strongly magnetised
($B\sim$10$^{15}$ G), slowly rotating ($P\sim$ 5-8 s) neutron
stars powered by the decay of the magnetic energy, rather than by
rotation (\cite{dt92}, \cite{pac92}, \cite{td95}).
They were discovered through
the detection of recurrent short ($\sim$0.1 s) bursts of
high-energy  radiation in the tens to hundreds of keV range, with
peak luminosity up to 10$^{39}$-10$^{42}$ erg s$^{-1}$, above the
Eddington limit for neutron stars. The rate of burst emission in
SGRs is highly variable. Bursts are generally emitted during
sporadic periods of activity, lasting days to months, followed by
long ``quiescent'' time intervals (up to years or decades) during
which no bursts are emitted. Occasionally SGRs emit also ``giant
flares'', that last up to a few hundred seconds
and have peak luminosity up to  10$^{46}$-10$^{47}$ erg s$^{-1}$.
Only three giant flares have been observed to date, each one from
a different source (see, e.g., \cite{mazets}, \cite{hurley1999},
\cite{swiftgiant}).

Persistent (i.e. non-bursting) emission is also observed from SGRs
in the soft X--ray range ($<$10 keV), with typical luminosity of
$\sim$10$^{35}$ erg s$^{-1}$, and, in three cases, periodic
pulsations at a few seconds have been detected. Such pulsations
proved the neutron star nature of SGRs and allowed to infer
spin-down at rates of $\sim$10$^{-10}$ s s$^{-1}$, consistent with
dipole radiation losses for magnetic fields of the order of
B$\sim$10$^{14}$-10$^{15}$ G. The X--ray spectra are generally
described with absorbed power laws, but in some cases strong
evidence has been found for the presence of an additional
blackbody-like component with typical temperature of $\sim$0.5 keV
(\cite{xmm}).

The only SGR for which persistent (i.e. not due to bursts and/or
flares) emission above 20 keV  has been reported to date is SGR
1806--20 (\cite{mereghetti05,molkov}).
Here we report the discovery, based on observations with the
\integral~ satellite (\cite{integral}), of persistent hard X-ray
emission from  \src~ in the 20-100 keV range.

\section{Observations and data analysis}

We  analysed data obtained with ISGRI (\cite{isgri}), the
low-energy  detector of the IBIS (\cite{ibis}) coded mask
telescope. IBIS/ISGRI is an imaging instrument covering a wide
field of view (29$^{\circ}\times$29$^{\circ}$ at zero sensitivity,
9$^{\circ}\times$9$^{\circ}$ at full sensitivity) with
unprecedented sensitivity and angular resolution
($\sim$12$^{\prime}$) in the hard X/soft $\gamma$-ray energy range
(15 keV-1 MeV). These excellent imaging performances are
essential, especially in crowded Galactic fields, to avoid source
confusion, which affected most previous experiments operating in
this energy range.

From the \integral~ public data archive we selected all the
observations pointed within 10$^{\circ}$ from the position of
\src. The resulting  data set consists of 1033 pointings, yielding
a total exposure time of $\sim$2.5 Ms. The observation period,
during which the source was observed discontinuously, spans from
March 6$^{th}$ 2003 to June 8$^{th}$ 2004. We used version 5.1 of
the Offline Scientific Analysis (OSA) Software provided by the
\integral~ Science Data Centre (\cite{isdc}).

\begin{figure}[ht!]
\centerline{\psfig{figure=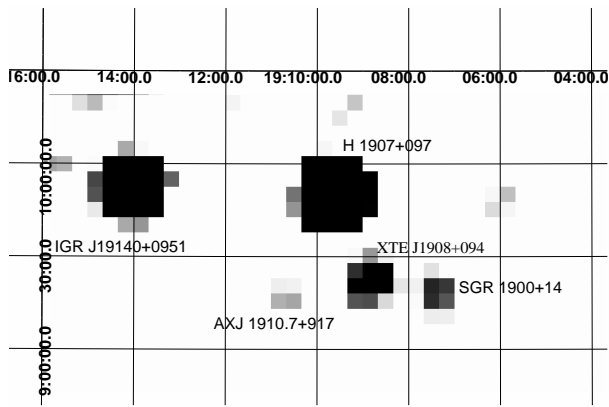,width=8cm}} \caption{IBIS/ISGRI
image of the \src~ field in the 18-60 keV energy range. The other
detected sources are the high-mass X-ray binary (HMXB) IGR
J19140+0951 (\cite{rodriguez}),  the black hole candidate XTE
J1908+094 (\cite{intz}), the HMXB pulsar H 1907+097
(\cite{makishima84}), and the weak unidentified source AXJ
1910.7+917 (\cite{sugizaki}).}

\label{img}
\end{figure}

After standard data processing (dead time correction, good
time-interval selection, gain correction, energy reconstruction),
we produced the sky images of each pointing in the 18--60 keV
range. These individual images were summed to produce a total
image, a portion of which is shown in Fig.~\ref{img}. A source
with count rate 0.18$\pm$0.02 counts s$^{-1}$ is detected with a
significance of 9$\sigma$ at coordinates (J2000) R.A. = 19$^{h}$
07$^{m}$ 25$^{s}$, Dec. = +09$^{\circ}$ 18$'$ 34$''$. The
associated error circle, with a radius of $\sim$3$^{\prime}$
(\cite{gros}) contains the position of \src~ (\cite{frail99}).  No
other  catalogued X-ray sources are present in the error circle.
We therefore associate the detected source with \src .

We found marginal evidence for a long term flux increase by
splitting the data in two parts (March to May 2003 and  November
2003 to June 2004) and producing two images of approximately equal
exposure. \src~ had a count rate of 0.14$\pm$0.03 counts s$^{-1}$
in the first period and of 0.25$\pm$0.03 counts s$^{-1}$ in the
second one.

To perform a spectral analysis we produced the summed images,
corresponding to the three time periods mentioned above, in five
energy bands (18-25, 25-35, 35-60, 60-100, and 100-200 keV). We
extracted the \src\ spectra using the source count rates obtained
from these images and rebinned the IBIS/ISGRI response matrix in
order to match our five energy channels. This spectral extraction
method, which we tested successfully with data also from the Crab
nebula, is particularly suited for weak sources that cannot be
detected in the individual pointings. Before fitting, we added a
systematic error of 5\% to the data, to account for the
uncertainties of our spectral extraction method and of the
response matrix.
In all cases a rather steep power law gave good
results (see Table \ref{sptab}). No signal was detected above 100
keV, but owing to the small statistics we could not establish the
possible presence of a spectral break at high energies.

\begin{table}[ht!]
\caption{Fluxes (20-100 keV) and spectral parameters derived for
\src~ with IBIS/ISGRI during the whole observation and during its
parts (see text). All the errors are at 1 $\sigma$ for one parameter of
interest.}
\begin{center}
\begin{tabular}{c|c|c}
 \hline\hline
 & Flux & Photon Index\\
 & [10$^{-12}$ erg cm$^{-2}$ s$^{-1}$] & \\
 \hline
 Average spectrum   & 15$\pm$3  & 3.1$\pm$0.5\\
 Spring 2003  & 10$\pm$3& 4.0$\pm$1.0 \\
 2003/2004 & 20$\pm$3 & 3.1$\pm$0.6\\
 \hline
\end{tabular}

\label{sptab}
\end{center}
\end{table}

\section{Discussion}

\src\ is the second SGR for which persistent hard X-ray emission
extending to $\sim$100 keV has been detected, the other being
\zerosei\ (\cite{mereghetti05,molkov}). The  spectrum of \src\ in
the 20-100 keV range, with photon index $\Gamma$=3.1$\pm$0.5, is
softer than that of \zerosei, which, in the last few years has
been the most active SGR.  In the latter source the photon index
varied from $\Gamma$=1.9$\pm$0.2, measured in the period March
2003-April 2004, to $\Gamma$=1.5$\pm$0.3 in September-October
2004, when the burst rate increased (\cite{mereghetti05}) before
the emission on December 27 2004 of the most powerful giant flare
ever observed from a SGR (Palmer et al. 2005, \cite{rhessigiant,acsgiant}).
Positive correlations between the bursting activity, the intensity
and hardness of the persistent emission, and the spin-down rate,
as have been observed in \zerosei\ (\cite{xmm}), are  expected in
magnetar models involving a twisted magnetosphere (\cite{tlk}),
since all these phenomena are driven by an increase of the twist
angle.

The soft spectrum of \src~ is possibly related to the fact that
this source is currently in a quiescent state. Short bursts were
observed from this source with BATSE (\cite{batse}), {\it RXTE} (\cite{rxte}) and
other satellites (e.g. \cite{sax,konus}) in the years 1979-2002. \src\ emitted a
giant flare on August 27 1998 (e.g. \cite{hurley1999}), followed by less intense
``intermediate'' flares on August 29 1998 (\cite{ibrahim}) and in April 2001 (\cite{guidorzi04,lenters}). The last bursts
reported from \src\ were observed with the Third Interplanetay
Network (IPN) in November 2002 (\cite{ipn1900}). No bursts from
this source were revealed in all the \integral~ observations from
2003 to 2005.

A comparison of the hard X-ray luminosity of the two SGRs is
subject to some uncertainties due to the unknown distances of
these sources. For \src\ a distance of 15 kpc has been derived
based on its likely association with a young star cluster
(\cite{vrba00}). For this distance the average flux of
about 1 mCrab corresponds to a 20-100 keV luminosity of
4$\times$10$^{35}$ erg s$^{-1}$. The distance of \zerosei\ is more
controversial. If also this source  is at $\sim$15 kpc
(\cite{MG05}), its hard X-ray luminosity would be at least three
times larger than that of \src. On the other hand, for a
distance in the 6.4 to 9.8 kpc range,  as derived from the latest
radio measurements of the afterglow of \zerosei~ giant flare
(\cite{cameron}), the two SGRs would have about the same
luminosity.

Hard X-ray persistent emission ($>$20 keV) has recently been
detected from another group of sources,  the Anomalous X-ray
Pulsars (AXPs, \cite{axps}), which share several
characteristics with the SGRs and are also believed to be
magnetars (see \cite{woodsrew}).
Hard X-ray emission has been detected from three AXPs with
\integral: 1E 1841--045 (\cite{molkovaxp}), 4U 0142+61
(\cite{denhartog}) and 1RXS J170849--400910 (\cite{revnitsev}).
The presence of pulsations seen with RXTE up to $\sim$200 keV in 1E
1841--045 (\cite{kuiper}) proofs that the hard X-ray emission
originates from the AXP and not from the associated supernova
remnant Kes 73. The discovery of (pulsed) persistent hard X-ray
tails in these three sources was quite unexpected, since below 10
keV the AXP have soft spectra, consisting of a blackbody-like
component (kT$\sim$0.5 keV) and a steep power law (photon index
$\sim$3--4).

\begin{figure}[ht!]
\centerline{\psfig{figure=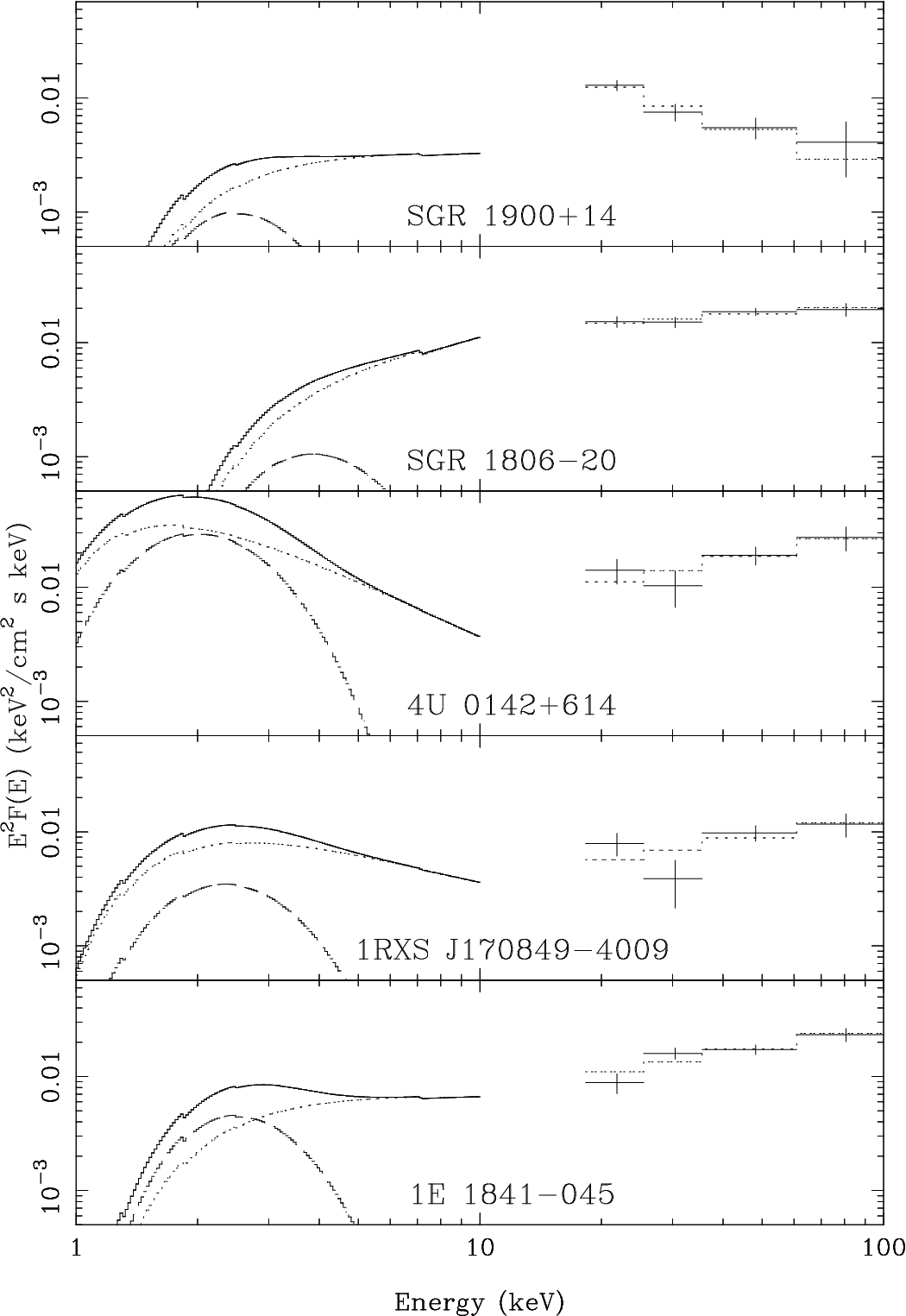,width=8cm}} 
\caption{Broad band X--ray spectra of the five magnetars detected by \integral. The
data points above 18 keV are the \integral~ spectra with their best fit
power-law models (dotted lines). The solid lines below 10 keV represent
the absorbed power-law (dotted lines) plus blackbody (dashed lines) models
taken from \cite{woods01} (\src, during a quiescent state in spring 2000), Mereghetti et al. (2005c) (\zerosei,
observation B, when the bursting activity was low), \cite{gohler} (4U~0142+614), \cite{rea}
(1RXS~J170849--4009), and \cite{morii} (1E~1841--045).}
\label{bbsp}
\end{figure}

In order to coherently compare the broad band spectral properties
of all the SGRs and AXPs detected at high energy, we analysed all
the public \integral~ data using the same procedures described
above for \src. Our results are summarised in Table \ref{magtab}
and in Fig. \ref{bbsp} where the \integral~ spectra are plotted together
with the results of observations at lower energy taken from the
literature (see figure caption for details). 
For \zerosei\ we
considered only the \integral~ data obtained from March to October
2003, in order to exclude the more active period observed in 2004.
We did not introduce any normalisation factor between the
different satellites and some discrepancy between the soft and
hard X-ray spectra might be ascribed to source variability since
the observations were not simultaneous. Nevertheless, even
considering these uncertainties, some indications  can be drawn
from the plotted spectra. Namely, in the three AXP a spectral
hardening above $\sim$10-20 keV is required (as already noted,
e.g. by Kuiper et al. (2004)), while 
at hard X/soft $\gamma$-rays the spectra of the two SGRs tend to be softer than the ones 
measured at low energies. The fact that the spectral break in \src~ is more
evident than in \zerosei~ could be also due to the different state during which the 
two sources have been observed, with the former in complete quiescence and the latter in
a low level of activity. All the three AXPs, on the other hand, can be considered in a quiescent
state since no SGR-like bursts have ever been reported from any of them.

\begin{table*}[ht!]
\caption{High-energy spectral parameters of Magnetars as measured by \integral.
The distances are taken from \cite{woodsrew} and references therein.}
\begin{center}
\begin{tabular}{c|c|c|c|c|c|c}
 \hline\hline
 Source & Exposure Time & Obs. Start & Obs. End & Photon Index & 20-100 keV Luminosity & Distance \\
 & [Ms] & UTC & UTC & & [10$^{34}$ erg s$^{-1}$] & [kpc] \\
 \hline
 SGR 1900+14& 2.5 & 2003-03-06 & 2004-06-08 & 3.1$\pm$0.5 & 40$\pm$8 & 15\\
 SGR 1806--20& 2.0 & 2003-03-12 & 2003-10-15 & 1.8$\pm$0.2 & 124$\pm$11 (35$\pm$3) & 15 (8)\\
 \hline
 4U 0142+61 & 0.33 & 2002-12-28 & 2004-06-09 & 1.3$\pm$0.4 & 5$\pm$0.5 &3\\
 1E 1841--045 & 1.5 & 2003-03-10 & 2004-05-02 & 1.5$\pm$0.2 & 26$\pm$2 & 7\\
 1RXS J170849.0--400910& 1.8 & 2003-02-01 & 2004-04-20 &1.4$\pm$0.4& 7$\pm$1 &5\\
 \hline
\end{tabular}
\label{magtab}
\end{center}
\end{table*}

\section{Conclusions}

We have discovered persistent emission in the 20-100 keV range
from \src. Its spectrum is softer than that of the only other SGRs
with persistent emission at these energies, \zerosei. This
difference is possibly due to the different activity state of the
two sources since \src~ was detected while it was quiescent from
the point of view of bursting emission, contrary to the case of
\zerosei.

Examining the broad band spectra of magnetars in the 1-100 keV
range, a notable difference between SGRs and AXPs appears. While
in SGRs the hard tails at higher energies are softer than the power law
components measured below 10 keV, in all the AXPs there is evidence for a 
spectral hardening between the soft and hard X-ray range.

In the framework of the magnetar model the persistent hard X-ray
emission can be powered either by bremsstrahlung photons produced
in a thin layer close to the neutron star surface, or at $\sim$100
km altitude in the magnetosphere through multiple resonant
cyclotron scattering (\cite{tlk,tb05}). The two models can be
distinguished by the presence of a cutoff at $\sim$ 100 keV or at
$\sim$1 MeV. Unfortunately the current \integral~ observations can
just firmly asses the presence of the high energy emission in
Magnetars, but cannot fully rule out the presence of spectral
breaks at high energies. Longer exposure times and/or observations
with more sensitive high-energy instruments are  required to
discriminate between the two models.

\begin{acknowledgements}
DG acknowledges support from the French Space Agency (CNES). ISGRI has been realized and maintained in flight
by CEA-Saclay/DAPNIA with the support of CNES. This work has been partially supported by the Italian Space Agency
and by the MIUR under grant PRIN 2004-023189.

\end{acknowledgements}

\end{document}